\documentclass[twocolumn, pra,showpacs,nofootinbib]{revtex4}
\usepackage{dcolumn,bm,graphicx}
\usepackage{amsmath}
\usepackage{amsfonts}
\usepackage{amssymb}

\begin{document}
\preprint{UNR Jan 2004-\today }
\title{Relaxation effect and radiative corrections in many-electron atoms}
\author{Andrei Derevianko}
\email{andrei@unr.edu}
\homepage{http://unr.edu/homepage/andrei}
\author{Boris Ravaine}
\affiliation {Department of Physics, University of Nevada, Reno,
Nevada 89557}
\author{W. R. Johnson}
\email{johnson@und.edu}
\homepage{http://www.nd.edu/~johnson}
\affiliation {Department of Physics,
University of Notre Dame, Notre Dame, Indiana 46556}

\date{\today}

\begin{abstract}
We illuminate the importance of a self-consistent many-body treatment
in calculations of vacuum polarization corrections to the energies of
atomic orbitals in many-electron atoms.
Including vacuum polarization in the
atomic Hamiltonian causes a substantial re-adjustment (relaxation) of
the electrostatic self-consistent field.
The induced change in the electrostatic energies is substantial for states with the
orbital angular momentum $l > 0$. For such orbitals, the relaxation mechanism
determines the sign and even the order of magnitude
of the total vacuum polarization correction.
This relaxation mechanism is illustrated with numerical results for the Cs atom.
\end{abstract}


\maketitle

Compared to hydrogenic one-electron systems, calculation
of radiative corrections for {\em many-electron}
atoms brings in an additional layer of complexity:
a strong Coulomb repulsion between the electrons.
The problem is especially challenging for
{\em neutral} many-electron atoms, where the interaction
of an outer-shell electron with other electrons is
comparable to its interaction with the nucleus.
At the same time, a reliable calculation of radiative corrections for
a heavy neutral system is
required in evaluation of the parity non-conserving (PNC) amplitude in
the 55-electron $^{133}$Cs atom.
Here it has been only recently realized that the sizes of radiative
corrections~\cite{Sus01,JohBedSof01,KucFla02,MilSusTer02,SapPacVei03} are
comparable to the experimental
error bar~\cite{WooBenCho97} of 0.35\% and, together with
the Breit correction~\cite{Der00},
dramatically affect agreement (or disagreement~\cite{BenWie99}) with
the Standard Model of elementary particles.

A systematic approach to the problem of radiative corrections
in strongly correlated systems is to start from a Furry representation based
on a self-consistent electronic potential~\cite{Sap98}. This
potential takes into account the fact that an
electron moves in an average field created by
both the nucleus and other electrons. Based on
this idea, a program of calculating
radiative corrections to PNC amplitudes
have been put forth by \citet{SapPacVei03}.
\citet{KucFla02} and \citet{MilSusTer02} pursue a more qualitative
approach using an independent-electron approximation.
We believe that the question of an interplay between
correlations and radiative corrections is yet to be
addressed. While here we do not compute the PNC corrections,
we illuminate a situation where disregarding correlations would
lead to a substantial error in determining radiative correction:
a radiative correction changes sign and even the order of
magnitude when the presence of other
electrons is accounted for.


In particular, we consider vacuum polarization (VP) corrections
to energies of atomic states. To the leading order in $\alpha Z$
the VP  may be  accounted for
by introducing the Uehling potential  $U_\text{VP}(r)$
into the atomic Hamiltonian. This potential is attractive,
and for a hydrogen-like ion the resulting VP corrections to the energies
are always {\em negative}.
For a complex atom, we find by contrast that, for
orbitals with  $l > 0$, the total correction is {\em positive}. Briefly, the reason
for such a counterintuitive effect is due to a
readjustment of atomic orbitals when the  $U_\text{VP}(r)$
potential is added to the self-consistent Dirac-Hartree-Fock (DHF)
equations.
The innermost $1s$ orbitals are ``pulled in'' by the short-ranged VP
potential, leading to a decrease of the effective nuclear
charge seen by the outer orbitals and thus to an increase
of the electrostatic energy of these orbitals.
Since for orbitals with  $l> 0$,  overlap with $U_\text{VP}(r)$
and thus the lowest order correction
are small, the resulting indirect ``relaxation'' contribution dominates
the total VP correction to the energies.
In the following we will present numerical results
supporting this relaxation mechanism.
Atomic units ($\hbar=|e|=m_e\equiv 1$) are used throughout.

Because of our interest in PNC in Cs, below we illustrate the relaxation effect with
numerical results for this atom; however,
the relaxation mechanism is also applicable in the cases of
other many-electron atoms. We also notice that the relaxation mechanism described
here is similar to that observed in calculations of the Breit
corrections~\cite{LinMarYnn89,Der02}.

The conventional many-electron Hamiltonian may be represented as
\begin{equation}
H= \sum_i h_0(i) + \frac{1}{2} \sum_{i \neq j} \frac{1}{r_{ij}} \, ,
\end{equation}
where the single-particle Dirac Hamiltonian is
\begin{equation}
 h_0(i) = c (\bm{\alpha}_i \cdot \bm{p}_i ) +
 \beta_i c^2 + V_\text{nuc}(r_i) \, .
\end{equation}
The nuclear potential $V_\text{nuc}(r)$ is obtained from
the nuclear charge distribution $\rho_\text{nuc}(r)$;
which is we approximate by the Fermi distribution
\begin{equation}
\rho_\text{nuc}(r)=\frac{\rho_{0}}{1+\exp[(r-c)/a]}\,,
\label{Eq:Fermi}
\end{equation}
where $\rho_{0}$ is the normalization constant, $c$ and $a$ are
the nuclear parameters. In our the numerical example for $^{133}$Cs,
we use $c=5.6748$~fm and $a=0.52$~fm.

A common starting point for describing a multi-electron atom
is the self-consistent field method. Here the many-body
wave-function is approximated by a Slater determinant
constructed from single-particle orbitals (bi-spinors) $u_k(\bm{r})$.
The orbitals are obtained by solving self-consistently
the eigenvalue equations
\begin{equation}
\left( h_0 + U_\text{DHF} \right) u_k(\bm{r}) =
\varepsilon_k u_k(\bm{r}) \, ,
\label{Eq:EqDHF}
\end{equation}
where $U_\text{DHF}$
is the traditional DHF potential which depends on the orbitals occupied
in the Slater determinant. The DHF energies for the core and several
valence orbitals of Cs are listed in Table~\ref{Tab:UehlingCorr}.

\begin{table}[h]
\caption{Vacuum polarization corrections to binding energies in neutral Cs ($Z=55$).
Here $\varepsilon_{nl_j}$ are the DHF energies, $\delta \varepsilon_{nl_j}^\text{(1)}$
are the expectation values of the Uehling potential (Eq.(\protect\ref{Eq:LowestOrder})),
and $\delta \varepsilon_{nl_j}^\text{DHF}$
are the VP corrections with the correlations included (Eq.(\protect\ref{Eq:DHFCorr})).
All quantities are
given in atomic units, $1 \, \text{a.u.} = 27.21138 \, \text{eV}$,
and notation $x[y]$ stands for $x \times 10^{y}$.
}
\label{Tab:UehlingCorr}
\begin{ruledtabular}
\begin{tabular}{cddd}
Orbital &
\multicolumn{1}{c}{$\varepsilon_{nl_j}$}&
\multicolumn{1}{c}{$\delta \varepsilon_{nl_j}^\text{(1)}$} &
\multicolumn{1}{c}{$\delta \varepsilon_{nl_j}^\text{DHF}$}  \\
\hline
\multicolumn{4}{c}{core orbitals}\\
$1s_{1/2}$ & -1330.396958  & -2.853[-1]   &   -2.782[-1]  \\
$2s_{1/2}$ &  -212.597116  & -3.392[-2]   &   -3.267[-2] \\
$2p_{1/2}$ &  -199.428898  & -1.510[-3]   &    5.406[-4] \\
$2p_{3/2}$ &  -186.434858  & -1.650[-4]   &    1.690[-3]  \\
$3s_{1/2}$ &   -45.976320  & -6.868[-3]   &   -6.581[-3]  \\
$3p_{1/2}$ &   -40.448097  & -3.339[-4]   &    1.987[-4]  \\
$3p_{3/2}$ &   -37.893840  & -3.719[-5]   &    4.609[-4]   \\
$3d_{3/2}$ &   -28.309043  & -1.839[-7]   &    4.531[-4]   \\
$3d_{5/2}$ &   -27.774710  & -4.370[-8]   &    4.425[-4]   \\
$4s_{1/2}$ &    -9.514218  & -1.457[-3]   &   -1.397[-3]   \\
$4p_{1/2}$ &    -7.446203  & -6.726[-5]   &    8.097[-5]    \\
$4p_{3/2}$ &    -6.920865  & -7.506[-6]   &    1.355[-4]   \\
$4d_{3/2}$ &    -3.485503  & -3.440[-8]   &    1.153[-4]    \\
$4d_{5/2}$ &    -3.396788  & -8.100[-9]   &    1.129[-4]   \\
$5s_{1/2}$ &    -1.490011  & -2.057[-4]   &   -2.050[-4]   \\
$5p_{1/2}$ &    -0.907878  & -7.773[-6]   &    2.035[-5]    \\
$5p_{3/2}$ &    -0.840312  & -8.395[-7]   &    2.757[-5]   \\
\multicolumn{4}{c}{valence  states}\\
$6s_{1/2}$ &   -0.127380   &  -1.054[-5]  &    -1.159[-5]     \\
$6p_{1/2}$ &   -0.085616   &  -1.942[-7]  &     2.284[-7]      \\
$6p_{3/2}$ &   -0.083785   &  -2.180[-8]  &     4.513[-7]     \\
$7s_{1/2}$ &   -0.055190   &  -2.896[-6]  &    -3.143[-6]    \\
$7p_{1/2}$ &   -0.042021   &  -6.957[-8]  &     8.150[-8]    \\
$7p_{3/2}$ &   -0.041368   &  -7.873[-9]  &     1.606[-7]   \\
\end{tabular}
\end{ruledtabular}
\end{table}

\begin{figure}[h]
\begin{center}
\includegraphics*[scale=0.7]{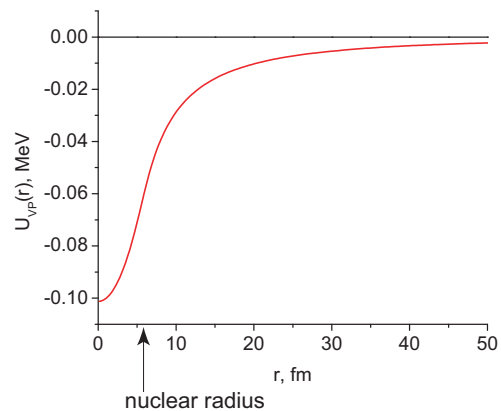}
\end{center}
\caption{Uehling potential for $^{133}$Cs. Notice that the radius of the innermost
$1s$ orbital is about $10^3$ fm, much larger than
the effective range of the VP potential.}
\label{Fig:UehCs}
\end{figure}

The polarization of the vacuum by the nucleus
modifies the nuclear electric field
seen by the electrons. To the leading order in $\alpha Z$, the VP may
be conveniently described with the Uehling potential, which
for a point-like nucleus of charge $Z$ reads
\begin{equation}
U^\text{p.c.}_\text{VP}\left(  r\right)  =\frac{2}{3\pi}\frac{\alpha Z}{r}\int_{1}^{\infty
}dt\sqrt{t^{2}-1}\left(  \frac{1}{t^{2}}+\frac{1}{2t^{4}}\right)  \exp\left[
-\frac{2r}{\alpha}t\right] \, .
\label{Eq:UehPot}
\end{equation}
This potential must be folded with the nuclear charge distribution,
\[
U_\text{VP}\left(  r\right)  = \int d \bm{r'}
\rho_\text{nuc}(|\bm{r} -\bm{r}' |) U^\text{p.c.}_\text{VP}\left( r' \right) \, .
\]
We approximated  $\rho_\text{nuc}(r)$ with
the Fermi distribution, Eq.~(\ref{Eq:Fermi}). In the numerical evaluation
of the extended-nucleus Uehling potential, we employed the routine
from Ref.~\cite{Hni94}. The Uehling potential $U_\text{VP}\left( r\right)$
generated by the Cs nucleus is shown in Fig.~\ref{Fig:UehCs}.
Notice that the actual range of this
potential is a few nuclear radii
(instead of Compton wavelength $\lambdabar_e \approx 384 \, \text{fm}$),
because the potential for a point-like charge, Eq.(\ref{Eq:UehPot}), diverges
logarithmically as $r \rightarrow 0$; therefore the folded potential $U_\text{VP}$
is dominated by the contributions accumulated inside the nucleus.

How does one compute the VP corrections $\delta \varepsilon_k$
to the energies of the atomic orbitals? Below we consider
two possibilities: (i) lowest-order perturbative treatment,
\begin{equation}
\delta \varepsilon_k^{(1)} = \langle u_k|U_\text{VP}| u_k\rangle \, ,
\label{Eq:LowestOrder}
\end{equation}
and (ii) the self-consistent approach. Indeed,
as in Ref.~\cite{JohBedSof01}, the VP potential may be introduced into the
DHF equations,
\begin{equation}
\left( h_0 + U_\text{VP} +U'_\text{DHF} \right) u'_k(\bm{r}) =
\varepsilon'_k u'_k(\bm{r}) \, ,
\label{Eq:EqUDHF}
\end{equation}
and a set of new energies $\varepsilon'_k$ and orbitals $u'_k(\bm{r})$
is obtained. Notice that the DHF potential is modified as well, since
it depends on the new set of the occupied orbitals $u'_k(\bm{r})$.
The {\em correlated} VP correction to the energy of the orbital $k$ is simply
\begin{equation}
\delta \varepsilon_{k}^\text{DHF} = \varepsilon'_k - \varepsilon_k \, .
\label{Eq:DHFCorr}
\end{equation}

Additionally, we carried out an independent correlated calculation in the
framework of the linearized Coupled DHF approximation \cite{Dal66},
which is equivalent to the random-phase approximation (RPA). This
approximation describes a linear response of the atomic orbitals to the
perturbing interaction, i.e. the VP potential. Numerical values
obtained from the linearized coupled DHF calculations were in close agreement
with the full DHF results.

The numerical results of our calculations are presented
in Table~\ref{Tab:UehlingCorr}.
While analyzing this Table, we observe that the lowest order corrections,
$\delta \varepsilon_k^{(1)}$, are always negative, reflecting
the fact that the Uehling potential is attractive (see Fig.~\ref{Fig:UehCs}).
Owing to the short-ranged nature of VP, and the fact that only  the $s$-orbitals
have a significant overlap with the nucleus,
the corrections to the energies of $l=0$ orbitals are much larger than
those for $l > 0$ orbitals. As to the correlated corrections, they differ quite
substantially from the lowest order-corrections. A comparison of
Eq.~(\ref{Eq:EqUDHF}) and Eq.~(\ref{Eq:EqDHF}) reveals the origin of this discrepancy:
the perturbation, in addition to the Uehling potential, contains a difference between the
two DHF potentials
\begin{equation}
\delta U = U_\text{VP} + \left( U'_\text{DHF} -  U_\text{DHF} \right) \, .
\end{equation}
For orbitals with $l>0$, where the first term above is small,
the modification of the DHF potential contributes significantly to the
VP energy corrections.

\begin{figure}[h]
\begin{center}
\includegraphics*[scale=1]{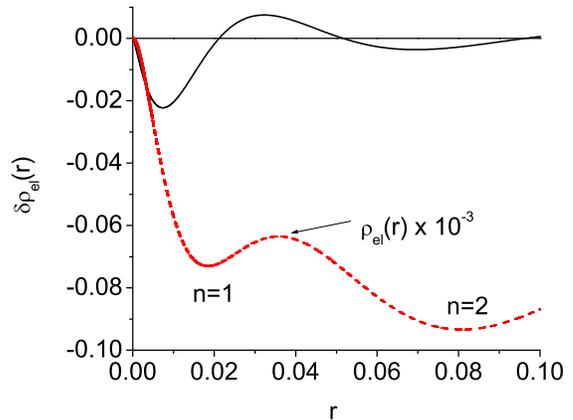}
\end{center}
\caption{Perturbation of the electronic radial charge distribution
$\delta \rho_\text{el} (r)$ (solid line)
 for Cs atom due to vacuum
polarization by the nucleus. We also show the unperturbed density
$\rho_\text{el} (r)$ multiplied by a factor of $10^{-3}$ (dashed line).
The minima of $\rho_\text{el} (r)$ correspond to positions of the
electronic shells, marked on the plot by their values of the principal quantum
number $n$.
 }
\label{Fig:rho}
\end{figure}

The modification of the DHF potential induced by the vacuum polarization
is clearly a many-body effect, not present in hydrogen-like system.
 Such an
effect has been explored before, for example in calculations of
the Breit corrections~\cite{LinMarYnn89,Der02},
and it is commonly referred to as a relaxation
mechanism. Let us illustrate this relaxation mechanism.
Denoting the correction to the occupied orbital wave functions as
$\chi_a(\bm r) = u'_a(\bm{r}) - u_a(\bm{r})$, we write
\begin{eqnarray*}
\lefteqn{ \left( U'_\text{DHF} -  U_\text{DHF} \right)  (\bm{r})\approx} \\
& &\sum_a  \int \chi^\dagger_a(\bm{r'})
\frac{1}{|r-r'|}
u_a(\bm{r'}) d\bm{r'} + \\
&&\sum_a \int u^\dagger_a(\bm{r'})
\frac{1}{|r-r'|}
\chi_a(\bm{r'}) d\bm{r'}-
\text{exchange} \, ,
\end{eqnarray*}
where we discarded contributions non-linear in $\chi_a(\bm r)$,
and ``exchange'' denotes non-local part of the perturbation.
The first two (direct) terms can be interpreted as an electrostatic
potential produced by a perturbation $\delta \rho_\text{el} (r)$
in the radial electronic density
\[
\rho_\text{el} (r) = - \frac{1}{4\pi r^2} \sum_a u^\dagger_a(\bm{r})u_a(\bm{r})\, .
\]
We plot both the electronic density $\rho_\text{el} (r)$ and
the VP-induced perturbation $\delta \rho_\text{el} (r)$
in Fig.~\ref{Fig:rho}. The minima of $\rho_\text{el} (r)$ correspond to
positions of the electronic shells, marked on the plot by their values of
principal quantum number $n$.

The figure~\ref{Fig:rho}  may be interpreted in the
following way:
the $s$ orbitals are ``pulled in'' by the attractive
Uehling potential closer to the nucleus. As a result,
 screening of the nuclear charge by the inner orbitals
becomes more efficient.  For example, the modification of the
effective charge felt by the $n=2$ electrons is simply the area under the
$\delta \rho_\text{el} (r)$ curve, accumulated between $r=0$
and the radius of the shell ($r \approx 0.08 a_0$);
from Fig.~\ref{Fig:rho} it is
clear that the induced modification of the effective charge for the  $n=2$ shell
has a negative sign.
Such an enhanced screening leads to a reduced attraction of  the electrons by the nucleus
and to the {\em increase} in the
energy of the outer electrons. From Table~\ref{Tab:UehlingCorr}, we see
that this indirect relaxation contribution to the energy may be
well comparable to the direct VP correction, $\delta \varepsilon^{(1)}_k$.
While for $l=0$ orbitals the direct correction gives a reasonable estimate,
for all orbitals with $l > 0$, the neglect of the relaxation would lead
to even qualitatively incorrect result.
Moreover, the higher the orbital angular momentum, the smaller is the direct
correction, and the more important is the relaxation mechanism.
For example, for $4d$ orbitals the VP correction in the lowest order is four orders
of magnitude smaller than the correlated result.

To summarize, here we illuminated the importance of the self-consistent many-body treatment
in calculations of vacuum polarization corrections.
Including the VP Uehling potential into the
atomic Hamiltonian causes re-adjustment (relaxation) of
the electrostatic self-consistent field.
The induced change in the electrostatic energies is substantial for states with the
orbital angular momentum $l > 0$. As illustrated in our numerical results for Cs,
the relaxation mechanism
determines the sign and even the order of magnitude
of the total VP correction for orbitals with $l > 0$.

\begin{acknowledgments}
The work of A. D. and B. R. was supported in part National Science Foundation
Grant No.\ PHY-00-99419,
the work of W.R.J. was supported in part National Science Foundation
Grant No.\ PHY-01-39928.
\end{acknowledgments}

\bibliography{general,exact,Breit,QED,pnc,mypub}


\end{document}